\documentclass[%
 reprint,
 amsmath,amssymb,
 aps,
]{revtex4-2}

\usepackage{graphicx}% Include figure files
\usepackage{dcolumn}% Align table columns on decimal point
\usepackage{bm}% bold math
\begin{document}

\preprint{APS/123-QED}

\title{Single-pixel diffuser camera}% Force line breaks with \\

\author{Baolei Liu}
\author{Fan Wang}
 \email{fan.wang@uts.edu.au}
\author{Chaohao Chen}
\author{David McGloin}
 \email{david.mcgloin@uts.edu.au}

\affiliation{%
 School of Electrical and Data Engineering, Faculty of Engineering and IT, University of Technology Sydney, NSW 2007, Australia
}%

\date{April 30, 2021}% It is always \today, today,

\begin{abstract}
We present a compact, diffuser assisted, single-pixel computational camera. A rotating ground glass diffuser is adopted, in preference to a commonly used digital micro-mirror device (DMD), to encode a two-dimensional (2D) image into single-pixel signals. We retrieve images with an 8.8\% sampling ratio after the calibration of the pseudo-random pattern of the diffuser under incoherent illumination. Furthermore, we demonstrate hyperspectral imaging with line array detection by adding a diffraction grating. The implementation results in a cost-effective single-pixel camera for high-dimensional imaging, with potential for imaging in non-visible wavebands.
\end{abstract}

\maketitle

%\tableofcontents

\section{Introduction} %%%%%%%%%%%% Introduction
By encoding depth or spectral information, imaging assisted by a diffuser or thin scatterer can retrieve a three-dimensional (3D) data cube rather than the conventional 2D image obtained from an optical sensor. The concept of a 'diffuser camera' ('DiffuserCam') \cite{antipa2018diffusercam} has gained a lot of interest, due to its compact and cost-effective approach. Retrieval of a three-dimensional (3D), multi-view \cite{zhu2019single}, multispectral \cite{sahoo2017single} or hyperspectral image \cite{monakhova2020spectral} via single-shot computational imaging has recently been demonstrated. However, such approaches are challenging for more esoteric wavelength bands, for instance, x-ray or terahertz imaging. As the name suggests, single-pixel imaging \cite{edgar2019principles,gibson2020single,zhang2015single,liu2017coloured,liu2018single}, produces images without the need for a 2D detector, making use of structured detection or illumination of the object to computationally derive an image. A such, single-pixel approaches are of interest in offering alternatives to conventional imaging, both for applications in the visible, but also as a low-cost alternative in regimes such as x-ray \cite{yu2016fourier,zhang2018tabletop}, infrared \cite{gibson2017real} and terahertz band \cite{olivieri2018time}, and even imaging atoms \cite{khakimov2016ghost}. Additionally, single-pixel approaches help to inform high-performance imaging techniques, for example, 3D depth, time-resolved or multispectral imaging, in which CCD based systems would be complicated or expensive to implement.  

Typically, a single-pixel camera uses a digital micromirror device (DMD), which is placed in the image plane, to modulate the image of objects with different 2D structured sampling patterns. The single-pixel detector then measures the corresponding total light intensity. By correlating the 1 dimensional (1D) single-pixel signals with the modulation patterns, reconstruction algorithms such as compressed sensing can rebuild the 2D image. Alternatively, the DMD can modulate the illumination of the object, an approach commonly called computational ghost imaging (CGI) \cite{shapiro2008computational,gibson2020single,liu2020self}. The DMD can be replaced by a liquid crystal spatial light modulator \cite{bromberg2009ghost}, rotating ground-glass diffusers \cite{valencia2005two} or LED arrays \cite{xu20181000,liu2016novel}. Image retrieval can be achieved using the same algorithms as in the structured detection scheme.

In cases where only passive imaging is needed, the structured detection scheme offers the benefit of a more compact and cheaper imaging system, due to the lack of a need for light sources.  However, when using wavelengths such as x-ray, conventional DMDs cannot be used for modulation. Thus, x-ray single-pixel imaging was realized using the CGI scheme, using, for example, a monochromatic x-ray beam passing through a slit array and a moving porous gold film \cite{yu2016fourier} or using polychromatic x-rays and a sheet of rotating sandpaper \cite{zhang2018tabletop} to generate pseudothermal illumination speckle patterns.

Here in this work, we present a single-pixel diffuser camera (SP-DiffuserCam) that uses a low-cost rotating ground glass diffuser instead of a DMD for 2D structured detection modulation with incoherent light. We show that the random and fixed patterns of a simple diffuser placed in the image plane can serve as 2D light intensity modulation in single-pixel imaging. We refer to this concept as a passive version of classical GI, but with no need for simultaneous measurements of reference patterns. We show this concept is readily extendable for hyperspectral imaging.

\section{Principle}
\label{sec:examples}
\begin{figure}[ht!]
\centering
\centering{\includegraphics[width=\linewidth]{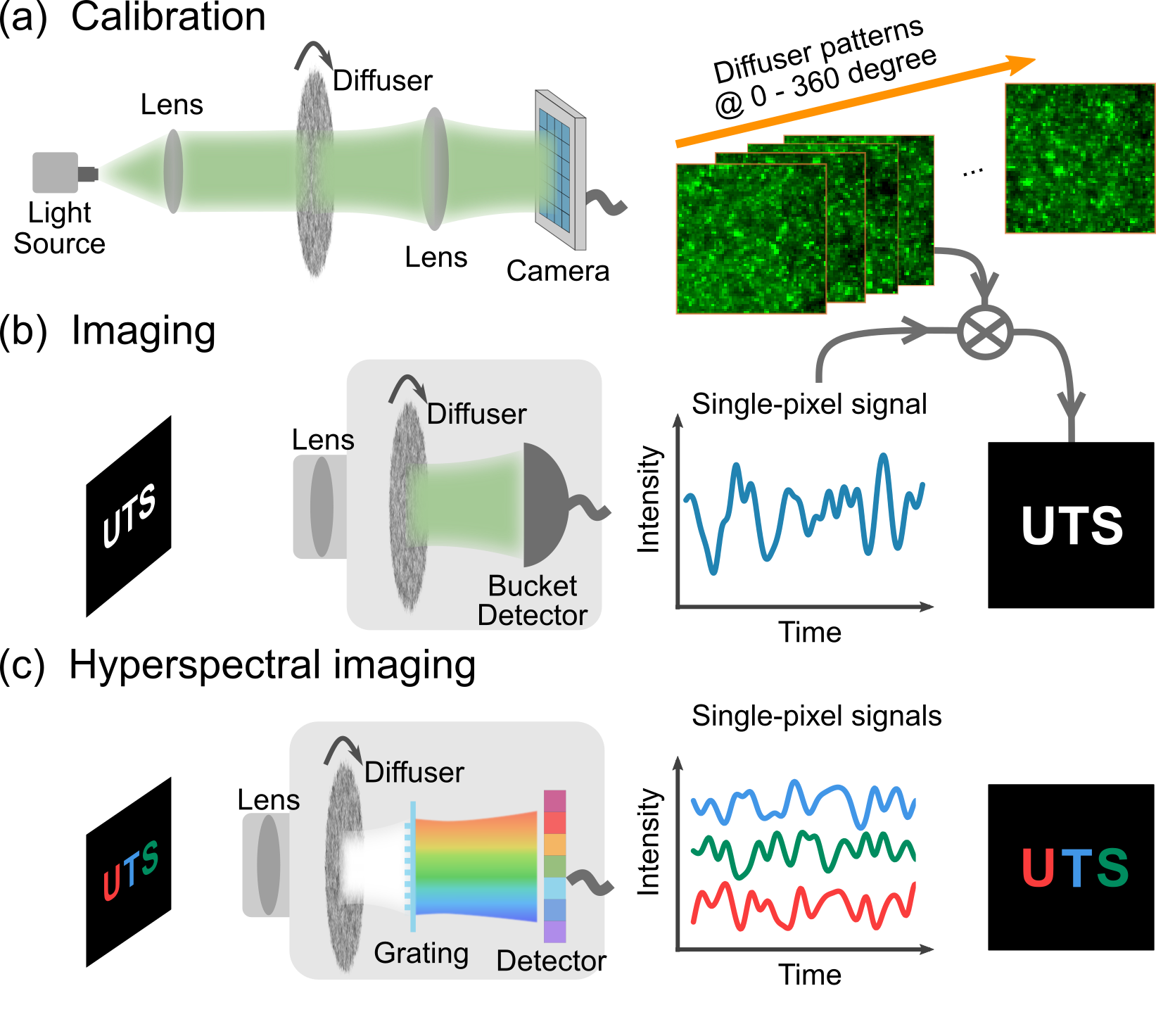}}
\caption{Principle of SP DiffuseCam. (a) Initially, a calibration procedure is conducted to acquire the structured patterns of the diffuser corresponding to different angular positions. (b) Once calibration is finished, the 2D image is retrieved by correlating the encoding patterns and the single-pixel temporal signal in repeated rotation, during which the object is imaged on the diffuser plane. (c) With a diffraction grating and line array detector, a hyperspectral imaging setup can be established with single-pixel signals from different wavelengths.}
\label{fig:1}
\end{figure}

Figure \ref{fig:1} presents the SP-DiffuserCam concept. The first procedure is a calibration process to map the speckle-liked patterns of the diffuser, where the intensity distributions $P(x,y,\theta)$ for each angle $\theta (0^{\circ}\leq\theta\leq360^{\circ})$ are acquired sequentially by rotating the diffuser. Note that this could be achieved using an laterally moving stage, but here we only consider rotation for system compactness. The second step is the temporal single-pixel measurement. An object $O(x,y)$ is illuminated by the same incoherent light source and imaged on the diffuser plane by a lens. The transmitted light intensity passing through the object $I(\theta)$, measured by a single-pixel detector under another repeated rotation, is simply the integration of the light on the detector plane,
\begin{equation}
I(\theta) = \int P(x,y,\theta)O(x,y)\, dx\,dy.
\label{eq:1}
\end{equation}
We then can acquire the object image $R(x,y)$ by using the differential correlation approach \cite{ferri2010differential},
\begin{equation}
R(x,y) = \left \langle I(\theta)P(x,y,\theta) \right \rangle - \frac{\left \langle I(\theta) \right \rangle}{\left \langle I_P(\theta) \right \rangle}\left \langle I_P(\theta)P(x,y,\theta) \right \rangle,
\label{eq:2}
\end{equation}
where $\left \langle \right \rangle$ denotes the ensemble average over the distribution of patterns and $I_P(\theta) = \int P(x,y,\theta)\, dx\,dy$ denotes the weights of the patterns.

For hyperspectral imaging, the detection intensity $I(\theta,\lambda)$ is simply the single-pixel signal corresponding to different wavelengths, measured using a line detector placed after a grating. Thus, the spectral image data cube can be reconstructed as
\begin{equation}
R(x,y,\lambda) = \left \langle I(\theta,\lambda)P(x,y,\theta) \right \rangle - \frac{\left \langle I(\theta,\lambda) \right \rangle}{\left \langle I_P(\theta) \right \rangle}\left \langle I_P(\theta)P(x,y,\theta) \right \rangle,
\label{eq:3}
\end{equation}
by only replacing the $I(\theta)$ with $I(\theta,\lambda)$ in Equation \ref{eq:2} and without characterizing the surface roughness for the other wavelengths. 

\section{Experiment} %Experimental result
\subsection{Single color SP-DiffuserCam}
\begin{figure}[ht!]
\centering
\centering\includegraphics[width=\linewidth]{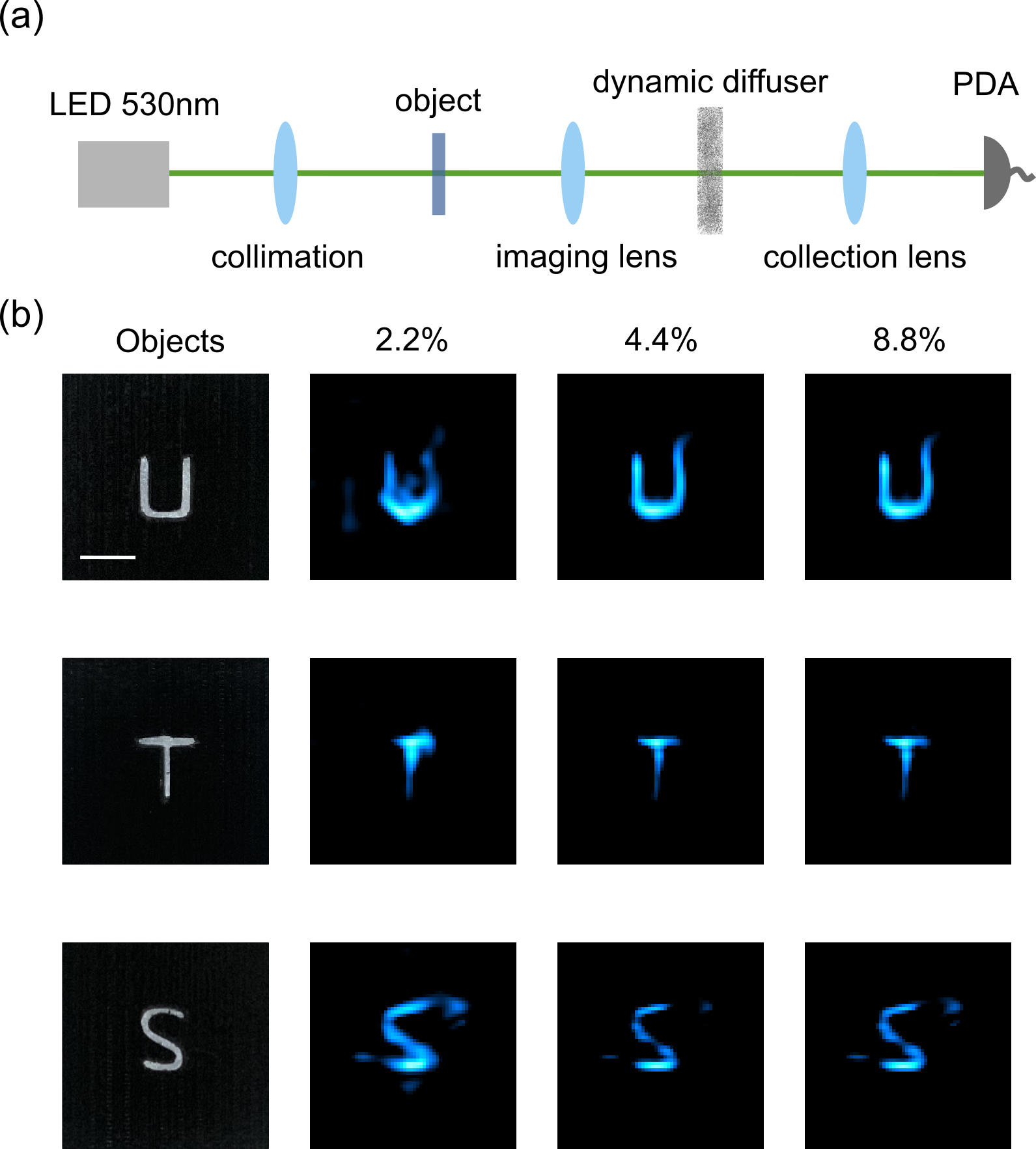}
\caption{Demonstration of SP-DiffuserCam for monochrome imaging. (a) Experimental setup. The object is imaged on the rotating diffuser plane, and then the temporal signal is recorded by a free-space amplified photodetector (PDA). (b) The retrieved images under increasing sampling ratios, compared to the original object (left column). Scale bar: 3mm.}
\label{fig:2}
\end{figure}

A simple proof-of-concept experimental realization of
the approach is presented in Figure \ref{fig:2}. Figure \ref{fig:2} (a) depicts the optical configuration of a single-color SP-DiffuserCam, where the light source is a monochromatic 530 nm LED with a bandwidth of 33 nm (M530D2, Thorlabs). The object is imaged on the diffuser plane, illuminated by the collimated light. The ground glass diffuser ($\varnothing$ = 24mm, DG10-120-MD, Thorlabs) is mounted on a motorized rotation stage (PRM1/MZ8, Thorlabs), which has a 25 $^{\circ}$/second rotation velocity. A silicon amplified photodetector (PDA100A2, Thorlabs) is used to measure the intensity fluctuations. For calibration, the speckle patterns $P(x,y,\theta)$ of the diffuser are recorded by a camera (panda 4.2 bi, PCO AG) through the same lens (f=50mm) used for the photodetector to keep the same numerical aperture; the exposure time of 4ms is the same for both detectors. A motor controller (KDC101, Thorlabs) and a low-cost DAQ device (USB-6002, National Instruments) are used to synchronize the rotator and the detector or the camera.

To ensure our patterns are sufficiently different, we use the edge of the diffuser region(about $d=6mm$ from the center of the diffuser) as the image plane area. Due to the small diameter of the diffuser, however, the neighbouring patterns still have similar areas along the rotation direction. The objects (Figure \ref{fig:2}(b) left) are 3D printed transmission masks of 'U', 'T', 'S' with a thickness of 2mm and the size about 3mm. The retrieved $64\times64$ pixel images with sampling ratios of 2.2\%, 4.4\%, and 8.8\% are shown in Figure \ref{fig:2}(b), corresponding to a minimum acquisition time of 14.4s, which is limited by the maximum rotation velocity. Here we also use a simple block-matching and the collaborative filtering algorithm \cite{dabov2007image} for noise suppression, which takes about 0.02 seconds. Outlines of objects emerge at 2.2\%, with image quality increasing with improved sampling ratios to 8.8\%. Further increasing the number of scattering patterns would not improve the reconstruction quality. The first reason is the 'imperfect' reference patterns with overlap regions between neighbouring ones. The second one is that a higher sampling ratio leads to a finer angle variation of measurement within one cycle of rotation, and increased overlap areas of the reference patterns. Thus, the above challenges induce a phenomenon that the spatial information along the vertical direction of rotation is reconstructed better than the parallel ones. This is evident in Figure \ref{fig:2}(b), in the top-right corner of 'U', the bottom of 'T' and the emerging noisy pixels in the left of 'S'. Note the rotation directions are clockwise the three masks, with the sampling area in the lower half of the circular diffuser.

\subsection{Comparison of different diffusers}

\begin{figure}[ht!]
\centering\includegraphics[width=\linewidth]{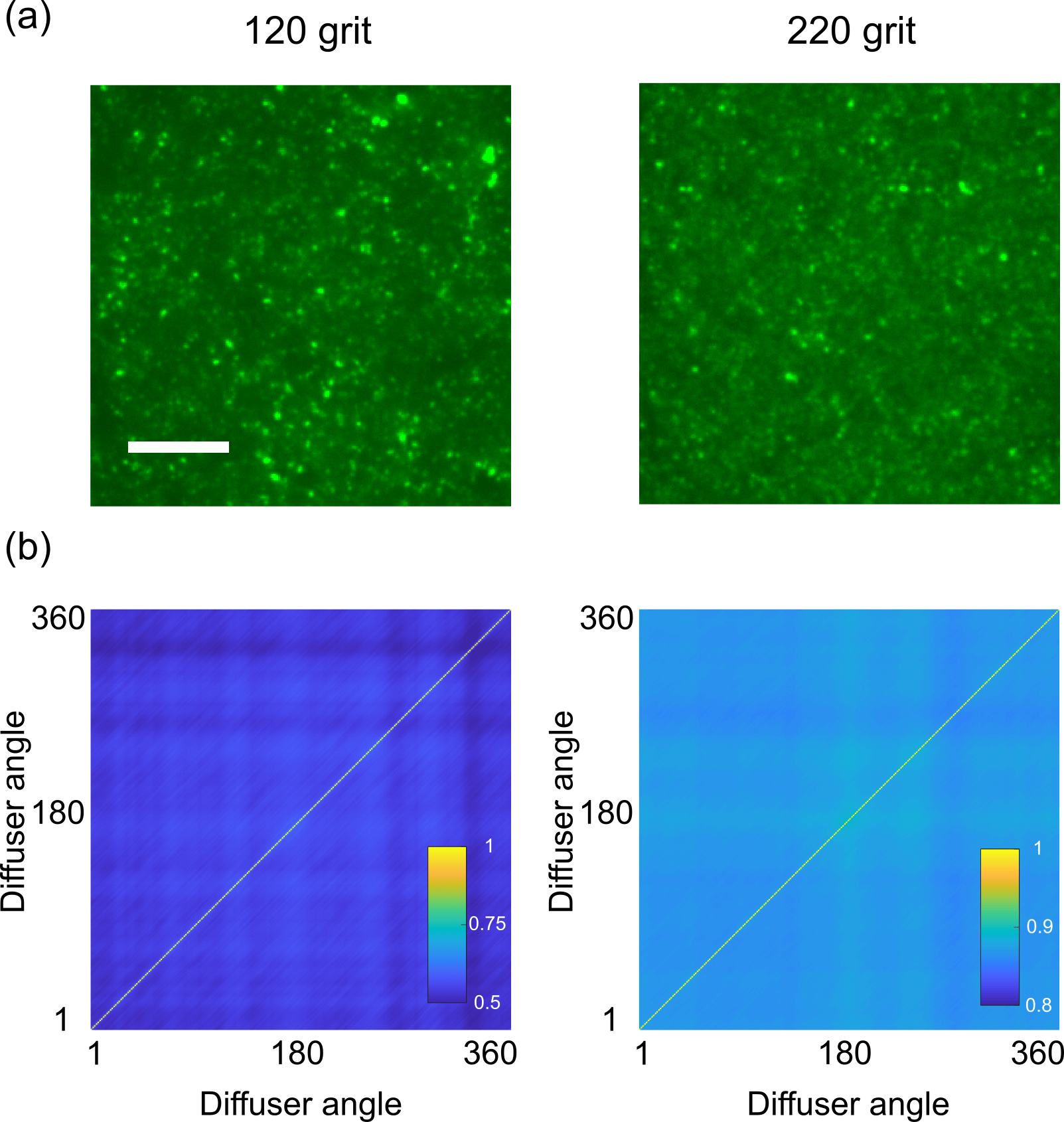}
\caption{Comparison of diffusers with coarse and fine scattering: (a) Example spatial images of 120 grit and 220 grit diffusers. Scale bar: 0.5mm. (b) The 2D correlation coefficients of corresponding two groups of diffuser patterns range from $\theta$ = 0$^{\circ}$,1$^{\circ}$,$\cdots$, 360$^{\circ}$ are depicted in the bottom row.}
\label{fig:3}
\end{figure}

The performance of our setup is largely dependent on the choice of diffuser. Here we compare the 120-grit diffuser, which is used in Figure \ref{fig:2}, and a 220-grit diffuser ($\varnothing$ = 24mm, DG10-220-MD, Thorlabs), which has smaller grain size on the polished surface. Figure \ref{fig:3}(a) shows typical transmission images of the two diffusers, in which the former one shows a higher contrast than the later one, due to the coarser polishing. We calculate the correlation coefficient $C(I_{\theta1},I_{\theta2})$ of the two groups of diffuser patterns, for a quantitative study, as defined by     
\begin{equation}
C(I_{\theta1},I_{\theta2}) = \frac{\int (I_{\theta1} - \bar{I}_{\theta1})(I_{\theta2} - \bar{I}_{\theta2})\, dx\,dy}{\sqrt{(\int (I_{\theta1} - \bar{I}_{\theta1})^2\,dx\,dy) (\int (I_{\theta2} - \bar{I}_{\theta2})^2\,dx\,dy})},
\label{eq:4}
\end{equation}
where $\theta$ = 0$^{\circ}$,1$^{\circ}$,$\cdots$, 360$^{\circ}$, $\bar{I}_{\theta1}$ and $\bar{I}_{\theta2}$ denote the average values for two arbitrary patterns $I_{\theta1}$ and $I_{\theta2}$, respectively. The mean correlation coefficients between different angles for the 120 grit diffuser is $C(I_{120}) = 0.55$, while for the 220 grit one we find $C(I_{220}) = 0.87$. Note that these coefficients are based on $1024\times1024$ pixels reference patterns, and they would be further higher in the condition of Figure \ref{fig:3}(b), where the patterns are binned to $64\times64$ matrices with lower contrasts. A higher correlation means an improved intrinsic coherence between the sampling basis and leads to a lower detection efficiency of the spatial images. Key to contemporary single-pixel imaging is the use of high-efficiency orthonormal patterns, such as Hadamard \cite{gibson2017real} or Fourier basis \cite{zhang2015single}. However, since we adopt a commercial diffuser, the sampling patterns in our setup are actually pseudo-random grayscale matrices, which are more related to classical GI using laser and dynamic diffuser induced speckle patterns \cite{valencia2005two}. According to a study of the influence of speckle size in GI \cite{sun2019toward}, an optimal speckle size exists in the range where the speckle size is comparable to the feature size of the object. This is the reason that we choose to use the 120-grit diffuser, as it is our coarsest diffuser and the one most approaches the feature size of the masks (about $0.4mm$). Future work could focus on the optimization of the diffuser or an integrated mini camera.    

\subsection{Hyperspectral SP-DiffuserCam}

\begin{figure}[ht!]
\centering\includegraphics[width=\linewidth]{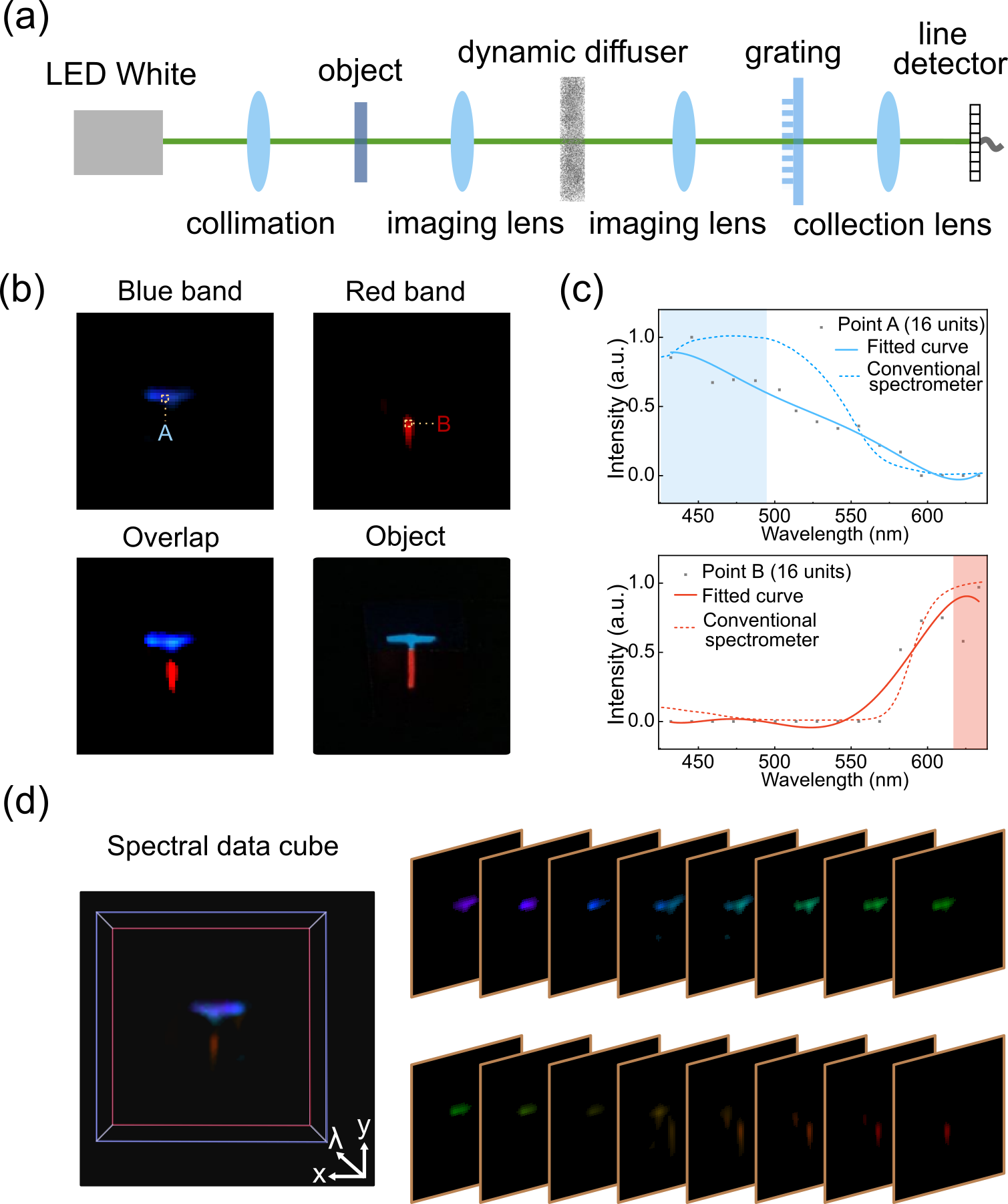}
\caption{Demonstration of SP-DiffuserCam for hyperspectral imaging. (a) Experimental setup. A white LED is used for illumination. A low-cost transmission grating is used for light dispersion after the structured modulation of a dynamic diffuser. (b) The retrieved images in the blue band ($461\pm35$ nm) and red band ($627\pm10$ nm) as well as their overlap, compared with the original object. (c) Reconstructed spectrum of points A and B in (b), compared with the conventional spectrometer measured ones. (d) The reconstructed spectral images and data cube. The pseudo-colored spectral images in (b) and (d) are converted from the spectra according to CIE color matching functions.}
\label{fig:4}
\end{figure}

One of the advantages of the SP-DiffuserCam is that it is easy to integrate other functions within the platform, such as the hyperspectral imaging shown in Figure \ref{fig:4}. To do this, we can directly use the pre-recorded reference patterns without additional calibration for specific wavelengths. Figure \ref{fig:4} (a) illustrates the optical setup of the hyperspectral SP-DiffuserCam. Here we use a cold white LED (MCWHLP1, Thorlabs) for a broad spectrum. Different from the monochrome setup, a low-cost diffraction grating slide (1,000 Line/mm, Rainbow Symphony, \$0.40), is used to disperse the incident light. The object image on the diffuser passes through the grating after the imaging lens (f=50mm) and is focused onto a line array detector via the collection lens (f=35mm). Here a compact COMS camera (acA1920-150um, Basler) serves as line-detector by binning the pixels to a $16\times1$ array in a sensor area of $9.5 mm\times3.3 mm$. To keep the same numerical aperture with the single-pixel measurement, we use the same two lenses system for calibration. %% pixels $1984\times174$ 

The 1D spectral intensity at each degree (exposure time: 4ms) are recorded in a 2D matrix $I(\theta,\lambda)$. Then this matrix is converted to a spectral data cube $R(x,y,\lambda)$ by the reconstruction Equation \ref{eq:3}. Here we apply standardized [International Commission on Illumination (CIE)] color-matching functions to the spectral cube to produce the pseudo-colored images to make consistent with the original object (made by sticking the color filters to the 'T' mask, a blue filter for the '-' and a red filter for the '$\vert$' of 'T'). Figure \ref{fig:4} (b) shows the retrieved spectral image in the blue band ($461\pm35 nm$) and the red band ($627\pm10nm$). Compared with the object photograph in the lower right corner, the middle, overlap area in the reconstruction has much lower intensities, due to refraction between the junction of the two filters. The reconstructed spectra of two selected areas in Figure \ref{fig:4} (b) show a similar trend when compared to measurements using a conventional spectrometer (UV-2401 PC UV-VIS, Shimadzu) of the same two color filters. Finally, we show the 3D spatio-spectral data cube in Figure \ref{fig:4} (d), as well as the spatial mapping image at single wavebands ranging from 426nm to 637 nm.

\section{conclusion} %%%%%%%%%%%%%% CONCLUSION
In conclusion, we have reported a passive single-pixel camera that employs a rotating diffuser as the spatial modulator in the image plane and uses incoherent light, for 2D imaging from 1D temporal signals. Note that classical pseudothermal ghost imaging  \cite{valencia2005two,shapiro2008computational,ferri2010differential,dou2020dark} has also used a rotating ground-glass diffuser illuminated by a laser beam to generate dynamic speckles, and then split the beam into a reference and an object beam. After that, a CCD is used to measure the diffracted speckles, and a single-pixel bucket detector is used to measure the signals from the object, simultaneously, in which both CCD and the object keep the same distances from the diffuser. As a comparison, our setup uses pre-recorded reference patterns and does not need to measure these in experiments. Another difference is the intensity distribution on the surface of the diffuser is used here instead of the laser interference speckles after the diffuser. In fact, our work can be considered as the passive version of x-ray ghost imaging that uses pre-recorded patterns and an incoherent source \cite{yu2016fourier,zhang2018tabletop}. However, we directly use the patterns of the diffuser as the reference patterns, not the diffracted speckles after propagation for a distance as in x-ray ghost imaging. Thus, our work would help to compact these imaging systems. We also demonstrate our concept is readily extended to achieve low-cost hyperspectral imaging, for 3D spatio-spectral image retrieval with temporally 1D spectral signals. Furthermore, the SP-DiffuserCam can be explored with coherent light, or other forms of imaging techniques such as time-resolved imaging \cite{sun2016single} using fast response detectors and phase imaging \cite{liu2018single,dou2020dark} by adopting phase engineered diffusers. We, therefore, anticipate that this work will open opportunities for developing cost-effective integrated single-pixel cameras, especially in exotic wavebands and imaging under ultra-low illumination.

\begin{acknowledgments}
Funding: Australian Research Council DECRA fellowship (DE200100074, F.W.), Australian Research Council Discovery Project (DP190101058, F.W.). The authors acknowledge financial support from the UTS Faculty of  Engineering and IT, and China Scholarship Council (B.L.: No.201706020170).
\end{acknowledgments}

\nocite{*}

\bibliography{SPDiffuserCam.bib}% Produces the bibliography via BibTeX.

\begin{thebibliography}{25}%
\makeatletter
\providecommand \@ifxundefined [1]{%
 \@ifx{#1\undefined}
}%
\providecommand \@ifnum [1]{%
 \ifnum #1\expandafter \@firstoftwo
 \else \expandafter \@secondoftwo
 \fi
}%
\providecommand \@ifx [1]{%
 \ifx #1\expandafter \@firstoftwo
 \else \expandafter \@secondoftwo
 \fi
}%
\providecommand \natexlab [1]{#1}%
\providecommand \enquote  [1]{``#1''}%
\providecommand \bibnamefont  [1]{#1}%
\providecommand \bibfnamefont [1]{#1}%
\providecommand \citenamefont [1]{#1}%
\providecommand \href@noop [0]{\@secondoftwo}%
\providecommand \href [0]{\begingroup \@sanitize@url \@href}%
\providecommand \@href[1]{\@@startlink{#1}\@@href}%
\providecommand \@@href[1]{\endgroup#1\@@endlink}%
\providecommand \@sanitize@url [0]{\catcode `\\12\catcode `\$12\catcode
  `\&12\catcode `\#12\catcode `\^12\catcode `\_12\catcode `\%12\relax}%
\providecommand \@@startlink[1]{}%
\providecommand \@@endlink[0]{}%
\providecommand \url  [0]{\begingroup\@sanitize@url \@url }%
\providecommand \@url [1]{\endgroup\@href {#1}{\urlprefix }}%
\providecommand \urlprefix  [0]{URL }%
\providecommand \Eprint [0]{\href }%
\providecommand \doibase [0]{https://doi.org/}%
\providecommand \selectlanguage [0]{\@gobble}%
\providecommand \bibinfo  [0]{\@secondoftwo}%
\providecommand \bibfield  [0]{\@secondoftwo}%
\providecommand \translation [1]{[#1]}%
\providecommand \BibitemOpen [0]{}%
\providecommand \bibitemStop [0]{}%
\providecommand \bibitemNoStop [0]{.\EOS\space}%
\providecommand \EOS [0]{\spacefactor3000\relax}%
\providecommand \BibitemShut  [1]{\csname bibitem#1\endcsname}%
\let\auto@bib@innerbib\@empty
%</preamble>
\bibitem [{\citenamefont {Antipa}\ \emph {et~al.}(2018)\citenamefont {Antipa},
  \citenamefont {Kuo}, \citenamefont {Heckel}, \citenamefont {Mildenhall},
  \citenamefont {Bostan}, \citenamefont {Ng},\ and\ \citenamefont
  {Waller}}]{antipa2018diffusercam}%
  \BibitemOpen
  \bibfield  {author} {\bibinfo {author} {\bibfnamefont {N.}~\bibnamefont
  {Antipa}}, \bibinfo {author} {\bibfnamefont {G.}~\bibnamefont {Kuo}},
  \bibinfo {author} {\bibfnamefont {R.}~\bibnamefont {Heckel}}, \bibinfo
  {author} {\bibfnamefont {B.}~\bibnamefont {Mildenhall}}, \bibinfo {author}
  {\bibfnamefont {E.}~\bibnamefont {Bostan}}, \bibinfo {author} {\bibfnamefont
  {R.}~\bibnamefont {Ng}},\ and\ \bibinfo {author} {\bibfnamefont
  {L.}~\bibnamefont {Waller}},\ }\bibfield  {title} {\bibinfo {title}
  {Diffusercam: lensless single-exposure 3d imaging},\ }\href@noop {}
  {\bibfield  {journal} {\bibinfo  {journal} {Optica}\ }\textbf {\bibinfo
  {volume} {5}},\ \bibinfo {pages} {1} (\bibinfo {year} {2018})}\BibitemShut
  {NoStop}%
\bibitem [{\citenamefont {Zhu}\ \emph {et~al.}(2019)\citenamefont {Zhu},
  \citenamefont {Sahoo}, \citenamefont {Wang}, \citenamefont {Lam},
  \citenamefont {Surman}, \citenamefont {Li},\ and\ \citenamefont
  {Dang}}]{zhu2019single}%
  \BibitemOpen
  \bibfield  {author} {\bibinfo {author} {\bibfnamefont {X.}~\bibnamefont
  {Zhu}}, \bibinfo {author} {\bibfnamefont {S.~K.}\ \bibnamefont {Sahoo}},
  \bibinfo {author} {\bibfnamefont {D.}~\bibnamefont {Wang}}, \bibinfo {author}
  {\bibfnamefont {H.~Q.}\ \bibnamefont {Lam}}, \bibinfo {author} {\bibfnamefont
  {P.~A.}\ \bibnamefont {Surman}}, \bibinfo {author} {\bibfnamefont
  {D.}~\bibnamefont {Li}},\ and\ \bibinfo {author} {\bibfnamefont
  {C.}~\bibnamefont {Dang}},\ }\bibfield  {title} {\bibinfo {title}
  {Single-shot multi-view imaging enabled by scattering lens},\ }\href@noop {}
  {\bibfield  {journal} {\bibinfo  {journal} {Optics Express}\ }\textbf
  {\bibinfo {volume} {27}},\ \bibinfo {pages} {37164} (\bibinfo {year}
  {2019})}\BibitemShut {NoStop}%
\bibitem [{\citenamefont {Sahoo}\ \emph {et~al.}(2017)\citenamefont {Sahoo},
  \citenamefont {Tang},\ and\ \citenamefont {Dang}}]{sahoo2017single}%
  \BibitemOpen
  \bibfield  {author} {\bibinfo {author} {\bibfnamefont {S.~K.}\ \bibnamefont
  {Sahoo}}, \bibinfo {author} {\bibfnamefont {D.}~\bibnamefont {Tang}},\ and\
  \bibinfo {author} {\bibfnamefont {C.}~\bibnamefont {Dang}},\ }\bibfield
  {title} {\bibinfo {title} {Single-shot multispectral imaging with a
  monochromatic camera},\ }\href@noop {} {\bibfield  {journal} {\bibinfo
  {journal} {Optica}\ }\textbf {\bibinfo {volume} {4}},\ \bibinfo {pages}
  {1209} (\bibinfo {year} {2017})}\BibitemShut {NoStop}%
\bibitem [{\citenamefont {Monakhova}\ \emph {et~al.}(2020)\citenamefont
  {Monakhova}, \citenamefont {Yanny}, \citenamefont {Aggarwal},\ and\
  \citenamefont {Waller}}]{monakhova2020spectral}%
  \BibitemOpen
  \bibfield  {author} {\bibinfo {author} {\bibfnamefont {K.}~\bibnamefont
  {Monakhova}}, \bibinfo {author} {\bibfnamefont {K.}~\bibnamefont {Yanny}},
  \bibinfo {author} {\bibfnamefont {N.}~\bibnamefont {Aggarwal}},\ and\
  \bibinfo {author} {\bibfnamefont {L.}~\bibnamefont {Waller}},\ }\bibfield
  {title} {\bibinfo {title} {Spectral diffusercam: Lensless snapshot
  hyperspectral imaging with a spectral filter array},\ }\href@noop {}
  {\bibfield  {journal} {\bibinfo  {journal} {Optica}\ }\textbf {\bibinfo
  {volume} {7}},\ \bibinfo {pages} {1298} (\bibinfo {year} {2020})}\BibitemShut
  {NoStop}%
\bibitem [{\citenamefont {Edgar}\ \emph {et~al.}(2019)\citenamefont {Edgar},
  \citenamefont {Gibson},\ and\ \citenamefont {Padgett}}]{edgar2019principles}%
  \BibitemOpen
  \bibfield  {author} {\bibinfo {author} {\bibfnamefont {M.~P.}\ \bibnamefont
  {Edgar}}, \bibinfo {author} {\bibfnamefont {G.~M.}\ \bibnamefont {Gibson}},\
  and\ \bibinfo {author} {\bibfnamefont {M.~J.}\ \bibnamefont {Padgett}},\
  }\bibfield  {title} {\bibinfo {title} {Principles and prospects for
  single-pixel imaging},\ }\href@noop {} {\bibfield  {journal} {\bibinfo
  {journal} {Nature Photonics}\ }\textbf {\bibinfo {volume} {13}},\ \bibinfo
  {pages} {13} (\bibinfo {year} {2019})}\BibitemShut {NoStop}%
\bibitem [{\citenamefont {Gibson}\ \emph {et~al.}(2020)\citenamefont {Gibson},
  \citenamefont {Johnson},\ and\ \citenamefont {Padgett}}]{gibson2020single}%
  \BibitemOpen
  \bibfield  {author} {\bibinfo {author} {\bibfnamefont {G.~M.}\ \bibnamefont
  {Gibson}}, \bibinfo {author} {\bibfnamefont {S.~D.}\ \bibnamefont
  {Johnson}},\ and\ \bibinfo {author} {\bibfnamefont {M.~J.}\ \bibnamefont
  {Padgett}},\ }\bibfield  {title} {\bibinfo {title} {Single-pixel imaging 12
  years on: a review},\ }\href@noop {} {\bibfield  {journal} {\bibinfo
  {journal} {Optics Express}\ }\textbf {\bibinfo {volume} {28}},\ \bibinfo
  {pages} {28190} (\bibinfo {year} {2020})}\BibitemShut {NoStop}%
\bibitem [{\citenamefont {Zhang}\ \emph {et~al.}(2015)\citenamefont {Zhang},
  \citenamefont {Ma},\ and\ \citenamefont {Zhong}}]{zhang2015single}%
  \BibitemOpen
  \bibfield  {author} {\bibinfo {author} {\bibfnamefont {Z.}~\bibnamefont
  {Zhang}}, \bibinfo {author} {\bibfnamefont {X.}~\bibnamefont {Ma}},\ and\
  \bibinfo {author} {\bibfnamefont {J.}~\bibnamefont {Zhong}},\ }\bibfield
  {title} {\bibinfo {title} {Single-pixel imaging by means of fourier spectrum
  acquisition},\ }\href@noop {} {\bibfield  {journal} {\bibinfo  {journal}
  {Nature Communications}\ }\textbf {\bibinfo {volume} {6}},\ \bibinfo {pages}
  {1} (\bibinfo {year} {2015})}\BibitemShut {NoStop}%
\bibitem [{\citenamefont {Liu}\ \emph {et~al.}(2017)\citenamefont {Liu},
  \citenamefont {Yang}, \citenamefont {Liu},\ and\ \citenamefont
  {Wu}}]{liu2017coloured}%
  \BibitemOpen
  \bibfield  {author} {\bibinfo {author} {\bibfnamefont {B.-L.}\ \bibnamefont
  {Liu}}, \bibinfo {author} {\bibfnamefont {Z.-H.}\ \bibnamefont {Yang}},
  \bibinfo {author} {\bibfnamefont {X.}~\bibnamefont {Liu}},\ and\ \bibinfo
  {author} {\bibfnamefont {L.-A.}\ \bibnamefont {Wu}},\ }\bibfield  {title}
  {\bibinfo {title} {Coloured computational imaging with single-pixel detectors
  based on a 2d discrete cosine transform},\ }\href@noop {} {\bibfield
  {journal} {\bibinfo  {journal} {Journal of Modern Optics}\ }\textbf {\bibinfo
  {volume} {64}},\ \bibinfo {pages} {259} (\bibinfo {year} {2017})}\BibitemShut
  {NoStop}%
\bibitem [{\citenamefont {Liu}\ \emph {et~al.}(2018)\citenamefont {Liu},
  \citenamefont {Suo}, \citenamefont {Zhang},\ and\ \citenamefont
  {Dai}}]{liu2018single}%
  \BibitemOpen
  \bibfield  {author} {\bibinfo {author} {\bibfnamefont {Y.}~\bibnamefont
  {Liu}}, \bibinfo {author} {\bibfnamefont {J.}~\bibnamefont {Suo}}, \bibinfo
  {author} {\bibfnamefont {Y.}~\bibnamefont {Zhang}},\ and\ \bibinfo {author}
  {\bibfnamefont {Q.}~\bibnamefont {Dai}},\ }\bibfield  {title} {\bibinfo
  {title} {Single-pixel phase and fluorescence microscope},\ }\href@noop {}
  {\bibfield  {journal} {\bibinfo  {journal} {Optics Express}\ }\textbf
  {\bibinfo {volume} {26}},\ \bibinfo {pages} {32451} (\bibinfo {year}
  {2018})}\BibitemShut {NoStop}%
\bibitem [{\citenamefont {Yu}\ \emph {et~al.}(2016)\citenamefont {Yu},
  \citenamefont {Lu}, \citenamefont {Han}, \citenamefont {Xie}, \citenamefont
  {Du}, \citenamefont {Xiao},\ and\ \citenamefont {Zhu}}]{yu2016fourier}%
  \BibitemOpen
  \bibfield  {author} {\bibinfo {author} {\bibfnamefont {H.}~\bibnamefont
  {Yu}}, \bibinfo {author} {\bibfnamefont {R.}~\bibnamefont {Lu}}, \bibinfo
  {author} {\bibfnamefont {S.}~\bibnamefont {Han}}, \bibinfo {author}
  {\bibfnamefont {H.}~\bibnamefont {Xie}}, \bibinfo {author} {\bibfnamefont
  {G.}~\bibnamefont {Du}}, \bibinfo {author} {\bibfnamefont {T.}~\bibnamefont
  {Xiao}},\ and\ \bibinfo {author} {\bibfnamefont {D.}~\bibnamefont {Zhu}},\
  }\bibfield  {title} {\bibinfo {title} {Fourier-transform ghost imaging with
  hard x rays},\ }\href@noop {} {\bibfield  {journal} {\bibinfo  {journal}
  {Physical Review Letters}\ }\textbf {\bibinfo {volume} {117}},\ \bibinfo
  {pages} {113901} (\bibinfo {year} {2016})}\BibitemShut {NoStop}%
\bibitem [{\citenamefont {Zhang}\ \emph {et~al.}(2018)\citenamefont {Zhang},
  \citenamefont {He}, \citenamefont {Wu}, \citenamefont {Chen},\ and\
  \citenamefont {Wang}}]{zhang2018tabletop}%
  \BibitemOpen
  \bibfield  {author} {\bibinfo {author} {\bibfnamefont {A.-X.}\ \bibnamefont
  {Zhang}}, \bibinfo {author} {\bibfnamefont {Y.-H.}\ \bibnamefont {He}},
  \bibinfo {author} {\bibfnamefont {L.-A.}\ \bibnamefont {Wu}}, \bibinfo
  {author} {\bibfnamefont {L.-M.}\ \bibnamefont {Chen}},\ and\ \bibinfo
  {author} {\bibfnamefont {B.-B.}\ \bibnamefont {Wang}},\ }\bibfield  {title}
  {\bibinfo {title} {Tabletop x-ray ghost imaging with ultra-low radiation},\
  }\href@noop {} {\bibfield  {journal} {\bibinfo  {journal} {Optica}\ }\textbf
  {\bibinfo {volume} {5}},\ \bibinfo {pages} {374} (\bibinfo {year}
  {2018})}\BibitemShut {NoStop}%
\bibitem [{\citenamefont {Gibson}\ \emph {et~al.}(2017)\citenamefont {Gibson},
  \citenamefont {Sun}, \citenamefont {Edgar}, \citenamefont {Phillips},
  \citenamefont {Hempler}, \citenamefont {Maker}, \citenamefont {Malcolm},\
  and\ \citenamefont {Padgett}}]{gibson2017real}%
  \BibitemOpen
  \bibfield  {author} {\bibinfo {author} {\bibfnamefont {G.~M.}\ \bibnamefont
  {Gibson}}, \bibinfo {author} {\bibfnamefont {B.}~\bibnamefont {Sun}},
  \bibinfo {author} {\bibfnamefont {M.~P.}\ \bibnamefont {Edgar}}, \bibinfo
  {author} {\bibfnamefont {D.~B.}\ \bibnamefont {Phillips}}, \bibinfo {author}
  {\bibfnamefont {N.}~\bibnamefont {Hempler}}, \bibinfo {author} {\bibfnamefont
  {G.~T.}\ \bibnamefont {Maker}}, \bibinfo {author} {\bibfnamefont {G.~P.}\
  \bibnamefont {Malcolm}},\ and\ \bibinfo {author} {\bibfnamefont {M.~J.}\
  \bibnamefont {Padgett}},\ }\bibfield  {title} {\bibinfo {title} {Real-time
  imaging of methane gas leaks using a single-pixel camera},\ }\href@noop {}
  {\bibfield  {journal} {\bibinfo  {journal} {Optics Express}\ }\textbf
  {\bibinfo {volume} {25}},\ \bibinfo {pages} {2998} (\bibinfo {year}
  {2017})}\BibitemShut {NoStop}%
\bibitem [{\citenamefont {Olivieri}\ \emph {et~al.}(2018)\citenamefont
  {Olivieri}, \citenamefont {Totero~Gongora}, \citenamefont {Pasquazi},\ and\
  \citenamefont {Peccianti}}]{olivieri2018time}%
  \BibitemOpen
  \bibfield  {author} {\bibinfo {author} {\bibfnamefont {L.}~\bibnamefont
  {Olivieri}}, \bibinfo {author} {\bibfnamefont {J.~S.}\ \bibnamefont
  {Totero~Gongora}}, \bibinfo {author} {\bibfnamefont {A.}~\bibnamefont
  {Pasquazi}},\ and\ \bibinfo {author} {\bibfnamefont {M.}~\bibnamefont
  {Peccianti}},\ }\bibfield  {title} {\bibinfo {title} {Time-resolved nonlinear
  ghost imaging},\ }\href@noop {} {\bibfield  {journal} {\bibinfo  {journal}
  {ACS Photonics}\ }\textbf {\bibinfo {volume} {5}},\ \bibinfo {pages} {3379}
  (\bibinfo {year} {2018})}\BibitemShut {NoStop}%
\bibitem [{\citenamefont {Khakimov}\ \emph {et~al.}(2016)\citenamefont
  {Khakimov}, \citenamefont {Henson}, \citenamefont {Shin}, \citenamefont
  {Hodgman}, \citenamefont {Dall}, \citenamefont {Baldwin},\ and\ \citenamefont
  {Truscott}}]{khakimov2016ghost}%
  \BibitemOpen
  \bibfield  {author} {\bibinfo {author} {\bibfnamefont {R.~I.}\ \bibnamefont
  {Khakimov}}, \bibinfo {author} {\bibfnamefont {B.}~\bibnamefont {Henson}},
  \bibinfo {author} {\bibfnamefont {D.}~\bibnamefont {Shin}}, \bibinfo {author}
  {\bibfnamefont {S.}~\bibnamefont {Hodgman}}, \bibinfo {author} {\bibfnamefont
  {R.}~\bibnamefont {Dall}}, \bibinfo {author} {\bibfnamefont {K.}~\bibnamefont
  {Baldwin}},\ and\ \bibinfo {author} {\bibfnamefont {A.}~\bibnamefont
  {Truscott}},\ }\bibfield  {title} {\bibinfo {title} {Ghost imaging with
  atoms},\ }\href@noop {} {\bibfield  {journal} {\bibinfo  {journal} {Nature}\
  }\textbf {\bibinfo {volume} {540}},\ \bibinfo {pages} {100} (\bibinfo {year}
  {2016})}\BibitemShut {NoStop}%
\bibitem [{\citenamefont {Shapiro}(2008)}]{shapiro2008computational}%
  \BibitemOpen
  \bibfield  {author} {\bibinfo {author} {\bibfnamefont {J.~H.}\ \bibnamefont
  {Shapiro}},\ }\bibfield  {title} {\bibinfo {title} {Computational ghost
  imaging},\ }\href@noop {} {\bibfield  {journal} {\bibinfo  {journal}
  {Physical Review A}\ }\textbf {\bibinfo {volume} {78}},\ \bibinfo {pages}
  {061802} (\bibinfo {year} {2008})}\BibitemShut {NoStop}%
\bibitem [{\citenamefont {Liu}\ \emph {et~al.}(2020)\citenamefont {Liu},
  \citenamefont {Wang}, \citenamefont {Chen}, \citenamefont {Dong},\ and\
  \citenamefont {McGloin}}]{liu2020self}%
  \BibitemOpen
  \bibfield  {author} {\bibinfo {author} {\bibfnamefont {B.}~\bibnamefont
  {Liu}}, \bibinfo {author} {\bibfnamefont {F.}~\bibnamefont {Wang}}, \bibinfo
  {author} {\bibfnamefont {C.}~\bibnamefont {Chen}}, \bibinfo {author}
  {\bibfnamefont {F.}~\bibnamefont {Dong}},\ and\ \bibinfo {author}
  {\bibfnamefont {D.}~\bibnamefont {McGloin}},\ }\bibfield  {title} {\bibinfo
  {title} {Self-evolving ghost imaging},\ }\href@noop {} {\bibfield  {journal}
  {\bibinfo  {journal} {arXiv preprint arXiv:2008.00648}\ } (\bibinfo {year}
  {2020})}\BibitemShut {NoStop}%
\bibitem [{\citenamefont {Bromberg}\ \emph {et~al.}(2009)\citenamefont
  {Bromberg}, \citenamefont {Katz},\ and\ \citenamefont
  {Silberberg}}]{bromberg2009ghost}%
  \BibitemOpen
  \bibfield  {author} {\bibinfo {author} {\bibfnamefont {Y.}~\bibnamefont
  {Bromberg}}, \bibinfo {author} {\bibfnamefont {O.}~\bibnamefont {Katz}},\
  and\ \bibinfo {author} {\bibfnamefont {Y.}~\bibnamefont {Silberberg}},\
  }\bibfield  {title} {\bibinfo {title} {Ghost imaging with a single
  detector},\ }\href@noop {} {\bibfield  {journal} {\bibinfo  {journal}
  {Physical Review A}\ }\textbf {\bibinfo {volume} {79}},\ \bibinfo {pages}
  {053840} (\bibinfo {year} {2009})}\BibitemShut {NoStop}%
\bibitem [{\citenamefont {Valencia}\ \emph {et~al.}(2005)\citenamefont
  {Valencia}, \citenamefont {Scarcelli}, \citenamefont {D’Angelo},\ and\
  \citenamefont {Shih}}]{valencia2005two}%
  \BibitemOpen
  \bibfield  {author} {\bibinfo {author} {\bibfnamefont {A.}~\bibnamefont
  {Valencia}}, \bibinfo {author} {\bibfnamefont {G.}~\bibnamefont {Scarcelli}},
  \bibinfo {author} {\bibfnamefont {M.}~\bibnamefont {D’Angelo}},\ and\
  \bibinfo {author} {\bibfnamefont {Y.}~\bibnamefont {Shih}},\ }\bibfield
  {title} {\bibinfo {title} {Two-photon imaging with thermal light},\
  }\href@noop {} {\bibfield  {journal} {\bibinfo  {journal} {Physical Review
  Letters}\ }\textbf {\bibinfo {volume} {94}},\ \bibinfo {pages} {063601}
  (\bibinfo {year} {2005})}\BibitemShut {NoStop}%
\bibitem [{\citenamefont {Xu}\ \emph {et~al.}(2018)\citenamefont {Xu},
  \citenamefont {Chen}, \citenamefont {Penuelas}, \citenamefont {Padgett},\
  and\ \citenamefont {Sun}}]{xu20181000}%
  \BibitemOpen
  \bibfield  {author} {\bibinfo {author} {\bibfnamefont {Z.-H.}\ \bibnamefont
  {Xu}}, \bibinfo {author} {\bibfnamefont {W.}~\bibnamefont {Chen}}, \bibinfo
  {author} {\bibfnamefont {J.}~\bibnamefont {Penuelas}}, \bibinfo {author}
  {\bibfnamefont {M.}~\bibnamefont {Padgett}},\ and\ \bibinfo {author}
  {\bibfnamefont {M.-J.}\ \bibnamefont {Sun}},\ }\bibfield  {title} {\bibinfo
  {title} {1000 fps computational ghost imaging using led-based structured
  illumination},\ }\href@noop {} {\bibfield  {journal} {\bibinfo  {journal}
  {Optics Express}\ }\textbf {\bibinfo {volume} {26}},\ \bibinfo {pages} {2427}
  (\bibinfo {year} {2018})}\BibitemShut {NoStop}%
\bibitem [{\citenamefont {Liu}\ \emph {et~al.}(2016)\citenamefont {Liu},
  \citenamefont {Yang}, \citenamefont {Zhang},\ and\ \citenamefont
  {Wu}}]{liu2016novel}%
  \BibitemOpen
  \bibfield  {author} {\bibinfo {author} {\bibfnamefont {B.-L.}\ \bibnamefont
  {Liu}}, \bibinfo {author} {\bibfnamefont {Z.-H.}\ \bibnamefont {Yang}},
  \bibinfo {author} {\bibfnamefont {A.-X.}\ \bibnamefont {Zhang}},\ and\
  \bibinfo {author} {\bibfnamefont {L.-A.}\ \bibnamefont {Wu}},\ }\bibfield
  {title} {\bibinfo {title} {A novel correlation imaging method using a
  periodic light source array},\ }\href@noop {} {\bibfield  {journal} {\bibinfo
   {journal} {Proc. SPIE}\ }\textbf {\bibinfo {volume} {10154}},\ \bibinfo
  {pages} {1015413} (\bibinfo {year} {2016})}\BibitemShut {NoStop}%
\bibitem [{\citenamefont {Ferri}\ \emph {et~al.}(2010)\citenamefont {Ferri},
  \citenamefont {Magatti}, \citenamefont {Lugiato},\ and\ \citenamefont
  {Gatti}}]{ferri2010differential}%
  \BibitemOpen
  \bibfield  {author} {\bibinfo {author} {\bibfnamefont {F.}~\bibnamefont
  {Ferri}}, \bibinfo {author} {\bibfnamefont {D.}~\bibnamefont {Magatti}},
  \bibinfo {author} {\bibfnamefont {L.}~\bibnamefont {Lugiato}},\ and\ \bibinfo
  {author} {\bibfnamefont {A.}~\bibnamefont {Gatti}},\ }\bibfield  {title}
  {\bibinfo {title} {Differential ghost imaging},\ }\href@noop {} {\bibfield
  {journal} {\bibinfo  {journal} {Physical Review Letters}\ }\textbf {\bibinfo
  {volume} {104}},\ \bibinfo {pages} {253603} (\bibinfo {year}
  {2010})}\BibitemShut {NoStop}%
\bibitem [{\citenamefont {Dabov}\ \emph {et~al.}(2007)\citenamefont {Dabov},
  \citenamefont {Foi}, \citenamefont {Katkovnik},\ and\ \citenamefont
  {Egiazarian}}]{dabov2007image}%
  \BibitemOpen
  \bibfield  {author} {\bibinfo {author} {\bibfnamefont {K.}~\bibnamefont
  {Dabov}}, \bibinfo {author} {\bibfnamefont {A.}~\bibnamefont {Foi}}, \bibinfo
  {author} {\bibfnamefont {V.}~\bibnamefont {Katkovnik}},\ and\ \bibinfo
  {author} {\bibfnamefont {K.}~\bibnamefont {Egiazarian}},\ }\bibfield  {title}
  {\bibinfo {title} {Image denoising by sparse 3-d transform-domain
  collaborative filtering},\ }\href@noop {} {\bibfield  {journal} {\bibinfo
  {journal} {IEEE Transactions on Image Processing}\ }\textbf {\bibinfo
  {volume} {16}},\ \bibinfo {pages} {2080} (\bibinfo {year}
  {2007})}\BibitemShut {NoStop}%
\bibitem [{\citenamefont {Sun}\ \emph {et~al.}(2019)\citenamefont {Sun},
  \citenamefont {Tuitje},\ and\ \citenamefont {Spielmann}}]{sun2019toward}%
  \BibitemOpen
  \bibfield  {author} {\bibinfo {author} {\bibfnamefont {Z.}~\bibnamefont
  {Sun}}, \bibinfo {author} {\bibfnamefont {F.}~\bibnamefont {Tuitje}},\ and\
  \bibinfo {author} {\bibfnamefont {C.}~\bibnamefont {Spielmann}},\ }\bibfield
  {title} {\bibinfo {title} {Toward high contrast and high-resolution
  microscopic ghost imaging},\ }\href@noop {} {\bibfield  {journal} {\bibinfo
  {journal} {Optics Express}\ }\textbf {\bibinfo {volume} {27}},\ \bibinfo
  {pages} {33652} (\bibinfo {year} {2019})}\BibitemShut {NoStop}%
\bibitem [{\citenamefont {Dou}\ \emph {et~al.}(2020)\citenamefont {Dou},
  \citenamefont {Cao}, \citenamefont {Gao},\ and\ \citenamefont
  {Song}}]{dou2020dark}%
  \BibitemOpen
  \bibfield  {author} {\bibinfo {author} {\bibfnamefont {L.-Y.}\ \bibnamefont
  {Dou}}, \bibinfo {author} {\bibfnamefont {D.-Z.}\ \bibnamefont {Cao}},
  \bibinfo {author} {\bibfnamefont {L.}~\bibnamefont {Gao}},\ and\ \bibinfo
  {author} {\bibfnamefont {X.-B.}\ \bibnamefont {Song}},\ }\bibfield  {title}
  {\bibinfo {title} {Dark-field ghost imaging},\ }\href@noop {} {\bibfield
  {journal} {\bibinfo  {journal} {Optics Express}\ }\textbf {\bibinfo {volume}
  {28}},\ \bibinfo {pages} {37167} (\bibinfo {year} {2020})}\BibitemShut
  {NoStop}%
\bibitem [{\citenamefont {Sun}\ \emph {et~al.}(2016)\citenamefont {Sun},
  \citenamefont {Edgar}, \citenamefont {Gibson}, \citenamefont {Sun},
  \citenamefont {Radwell}, \citenamefont {Lamb},\ and\ \citenamefont
  {Padgett}}]{sun2016single}%
  \BibitemOpen
  \bibfield  {author} {\bibinfo {author} {\bibfnamefont {M.-J.}\ \bibnamefont
  {Sun}}, \bibinfo {author} {\bibfnamefont {M.~P.}\ \bibnamefont {Edgar}},
  \bibinfo {author} {\bibfnamefont {G.~M.}\ \bibnamefont {Gibson}}, \bibinfo
  {author} {\bibfnamefont {B.}~\bibnamefont {Sun}}, \bibinfo {author}
  {\bibfnamefont {N.}~\bibnamefont {Radwell}}, \bibinfo {author} {\bibfnamefont
  {R.}~\bibnamefont {Lamb}},\ and\ \bibinfo {author} {\bibfnamefont {M.~J.}\
  \bibnamefont {Padgett}},\ }\bibfield  {title} {\bibinfo {title} {Single-pixel
  three-dimensional imaging with time-based depth resolution},\ }\href@noop {}
  {\bibfield  {journal} {\bibinfo  {journal} {Nature Communications}\ }\textbf
  {\bibinfo {volume} {7}},\ \bibinfo {pages} {1} (\bibinfo {year}
  {2016})}\BibitemShut {NoStop}%
\end{thebibliography}%

\end{document}